# Infrared properties of micromachined vanadium oxide thin films

Martin Rees, Haifei Wang, Thomas Decker, Robinson L. Smith

*Rio Salado College, Tempe, AZ, USA*

**ABSTRACT**

*This paper discusses questions of synthesizing and pressing vanadium oxides to create film-forming materials that can be used in producing optical coatings. Based on the film-forming materials thus created, technological processes have been developed for fabricating coatings from vanadium dioxide by two methods of producing thin films: vacuum evaporation and magnetron sputtering. Questions of using films made from vanadium oxide in optical instrumentation are considered.*

Low cost, reliable, uncooled JR detector arrays have potential for use in a variety of commercial applications, including transportation, security/law enforcement, industrial machine and process control, waste and pollutant detection, and medical imaging, as well as in military applications such as night vision equipment, missile seeker sensors, precision guided munitions, intrusion alarms and reconnaissance. Compared with the highest performing cryogenically cooled detectors, room temperature uncooled detectors including bolometers and pyroelectric devices, offer considerable advantages in cost and operational convenience, with minimal sacrifice in performance. In particular, some of the most important advantages of uncooled sensors also include: lower unit andlife cycle cost, reduced power consumption, smaller size and reduced weight, and finally potentially higher reliability. A limited number of prototype microbolometer arrays have been described in the literature. One of the latest disclosures is a 327 x 245 pixel array with integrated readout circuit fabricated at Loral Infrared & Imaging Systems1. Despite an evident lack of commercially available devices, predictions concerning the bolometric array market are quite encouraging. In this paper, 1 x 64, 1 x 128 and 1 x 256 pixel linear microbolometric detector arrays are presented. Their structure, fabrication process and performance are described.

A leading position among substances that can be used in the thin-film technology of optical coatings for the UV, visible, and near-IR regions is occupied by the oxides $ZrO_2$, $Y_2O_3$, $CeO_2$, $Al_2O_3$, etc., layers of which are transparent in a wide spectral range. The vanadium oxides, which possess specific thermochromic properties, are of great interest among the metal oxides.

Table 1: Thermodymanic properties of VO2 used here

| Reaction | $\Delta_r H_0^0$, kCal/mol |
|---|---|
| $2V_2O_5(sol) = V_4O_{10}(gas)$ | $72.71 \pm 2.0$ |
| $4V(sol) + 5O_2(gas) = V_4O_{10}(gas)$ | $-669.29 \pm 6.0$ |
| $2V_2O_5(sol) = V_4O_8(gas) + O_2(gas)$ | $154.29 \pm 3.0$ |
| $4V(sol) + VO_2(gas) = V_4O_8(gas)$ | $-587.71 \pm 6.0$ |
| $V_2O_3(sol) = VO_2(gas) + VO(gas)$ | $264.96 \pm 3.0$ |
| $VO(sol) = VO(gas)$ | $135.97 \pm 2.0$ |

Vanadium dioxide layers are attracting special attention as the basis of a new type of bistable controllable thin-film structures that are promising for developing optical information-transport and -processing systems, controlling the energy characteristics of optical beams, and the spatiotemporal correction of the parameters of laser radiation, as well as for creating elements to protect fiber-optic communication lines and optoelectronic devices from the action of powerful optical fields. It should be pointed out that the term "bistability" with respect to such structures is used here in a somewhat extended sense, referring both to the bistability proper that manifests itself during a thermochromic phase transition and to the presence of two stable states with sharply different optical properties in the structures. Electronic and optical components based on vanadium dioxide have already come into practical use, and these can include critical thermistors, optical switches, modulators, and controllable laser mirrors, as well as energy-conserving optical coatings for window glasses.2–5 The changes of the optical characteristics in crystalline VO2 are caused by the semiconductor-metal phase transition PT that occurs at a temperature of 69 °C. The amplitude of the indicated changes in similar VO2 films is largely determined by the physicochemical characteristics of the resulting layers. Improving the bistable properties of vanadium dioxide requires broadening the spectral range toward shorter wavelengths 0.6–2.5 μm, reducing the phase-transition temperature, and changing its slope and other characteristics. One way to achieve these goals is to modify the properties of bistable layers of VO2 by doping them with additional impurities of related compounds. In particular, this can conveniently be done by adding tungsten oxides and by varying the V1–xWxOy composition in a wide interval of x and y values. As is well known, vanadium gives compounds corresponding to valences 2, 3, 4, and 5, and mainly forms four oxides: VO, V2O3, VO2, and V2O5. The vanadium oxides of higher degrees of oxidation are in this case more important in practice, being more stable. However, besides these oxides, which correspond to all possible oxidation levels of vanadium, it forms a whole series of intermediate oxides, such as V3O7, V4O7, V6O13, V12O26, and V12O29, which are mainly solid solutions with a large number of vacancies in the anion lattice for example, their number sometimes reaches 30 in V12O29. 6 According to the data of Ref. 7, there are several low oxides of vanadium between VO2 and V2O3 whose composition can be expressed by the general formula VnO2n−1, where n=3–8. These oxides have very narrow regions of homogeneity. They are obtained by heating mixtures of V, V2O3, and V2O5 in vacuum at 650–1000 °C for 2–20 days.

Table 2: Thermodynamic data of the evaporation process of vanadium oxide.

| | | | | $\log P(\text{Torr}) = -(A/T) + B$ | |
|---|---|---|---|---|---|
| Reaction | $\Delta_r H_T^{0a}$ | $\Delta_r H_0^{0a}$ | $\Delta_r H_0^{0b}$ | $A$ | $B$ |
| $V_2O_3(\text{sol}) = VO(\text{gas}) + VO_2(\text{gas})$ | $256 \pm 3$ | $276 \pm 3$ | $274 \pm 1$ | 27900 | 11.34 |
| $V_2O_{3.04}(\text{sol}) = yVO_2(\text{gas}) + (x-y)VO(\text{gas})$ | — | — | $267 \pm 2$ | 27470 | 11.22 (VO₂) |
| $+ (2.04x-y)O(\text{gas})$ | | | | 28730 | 11.74 (VO) |
| $VO_2(\text{sol}) = VO_2(\text{gas})$ | 105 | — | $114 \pm 2$ | 22860 | 11.24 |
| $VO_2(\text{liq}) = VO_2(\text{gas})$ | — | — | — | 19861 | 9.45 |
| $2V_2O_5(\text{liq}) = V_4O_{10}(\text{gas})$ | $27 \pm 0.5$ | $71.5 \pm 0.5$ | $69.4 \pm 0.2$ | 5905 | 2.52 |

[a]Calculated from the second law of thermodynamics.
[b]Calculated from the third law of thermodynamics.

Microbolometers, which belong to the class of thermal detectors, consist of a thermal mass suspended by supports of low thermal conductance. Infrared radiation heats the suspended thermal mass, and the resulting temperature increase is measured by an embedded thermistor. This change is then converted into a voltage or current signal. The fabricated bolometric detectors are schematically shown in Fig. 1. The pixel thermistor material is a polycrystalline film of vanadium dioxide3. This material shows a temperature induced crystallographic transformation that is accompanied by a reversible semiconductor (low-temperature phase)-to-metal (high-temperature phase) phase transition, with a significant change in electrical and optical properties. V02 undergoes this transition typically at a temperature approaching 70°C. During the semiconductor-to-metal phase transition, the crystallographic structure of vanadium dioxide changes from monoclinic to tetragonal. An accompanying resistivity change by a factor as great as iO has been reported in single crystals. Resistivity change in thin films of V02 is usually less pronounced (102 1O) and depends on fabrication conditions and substrate material used. With a band gap energy of 0.6 to 0.7 eV, the semiconductor phase of V02 is highly transparent in the 3 to 12 pmregion of the spectrum. In the metallic phase, V02 has a high absorption coefficient and is strongly reflective.

Table 3: Properties of the VO2 samples after annealing.

| Annealing temperature, °C | Annealing time, h | Shrinkage in diameter, % | Shrinkage in height, % | Apparent density, g/cm$^3$ | Open porosity, % |
|---|---|---|---|---|---|
| 400 | 2 | 0 | 0 | 1.72 | 47.1 |
| 500 | 2 | 0 | 0 | 1.71 | 46.6 |
| 600 | 2 | 2.6 | 2.4 | 1.89 | 41.7 |

The V02 film fabrication method developed at the Institute allows us to vary the electrical properties of the films including their TCR. This is illustrated in Fig. 2. Routinely obtained TCR values are close to -3 % per degree centigrade but under certain conditions, TCR values up to -5 % per degree centigrade, can also be obtained. As was already mentioned, the JR radiation absorption in the V02-based pixels of the bolometric detectors operated near room temperature is relatively low. Since the presented bolometric detectors lack a dedicated absorber, the absorption spectrum is determined by the optical properties of the detector construction materials and thus can be tailored by appropriate choice of these thin film materials. A typical JR absorption spectrum at room temperature of a bolometric detector sandwich is shown in Fig. 3. Non-uniform absorbance of the detector, with a value near 50 % close to 10 p.m wavelength, is clearly visible. The bolometric detectors presented in this paper were fabricated using bulk silicon micromachining4. The fabrication process flow is schematically shown in Fig. 4. The thickness of the V02 and Si3N4 films were varied from 0. 1 to 0.5 pm. It should be mentioned that the described fabrication process is much simpler, i.e. consists of fewer steps, than the processes typically involved in surface micromachining of silicon4. On the other hand, the creation of pits in the silicon underneath the detector pixels excludes the possibility of fabrication of the readout electronics in this location. This makes the described technology quite suitable for fabrication of multipixel linear detector arrays but not so practical for fabrication of bi-dimensional arrays. In linear arrays, the readout electronics can be fabricated on the same silicon wafer adjacent to the detectors, while in bi-dimensional arrays, it would be preferable to place at least a part of the readout electronics directly underneath the detector pixels.

Scanning electron microscope (SEM) pictures of various fabricated bolometric detectors are shown in Fig. 4. These test detector structures consist of a V02 pixel supported by a micromembrane suspended by two or four arms of various shapes and sizes, over pits anisotropically etched in the silicon substrate. The number and dimensions of the supporting arms influence the sensitivity and response time of the detectors. The geometries with a limited number of long and narrow arms lead to an increase in detector sensitivity, but at the same time cause an increase in detector response time. For testing, the fabricated detectors are mounted in standard integrated circuit packages equipped with temperature stabilizing thermoelectric (TE) coolers. An example of a packaged die consisting of two 64-pixel linear detector arrays is shown in Fig. 3.

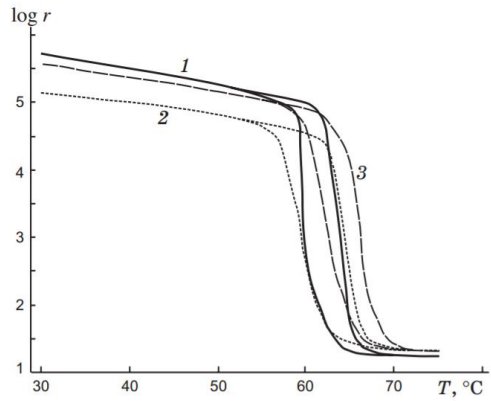

Figure 1: Temperature dependence of the resistance of the films referenced.

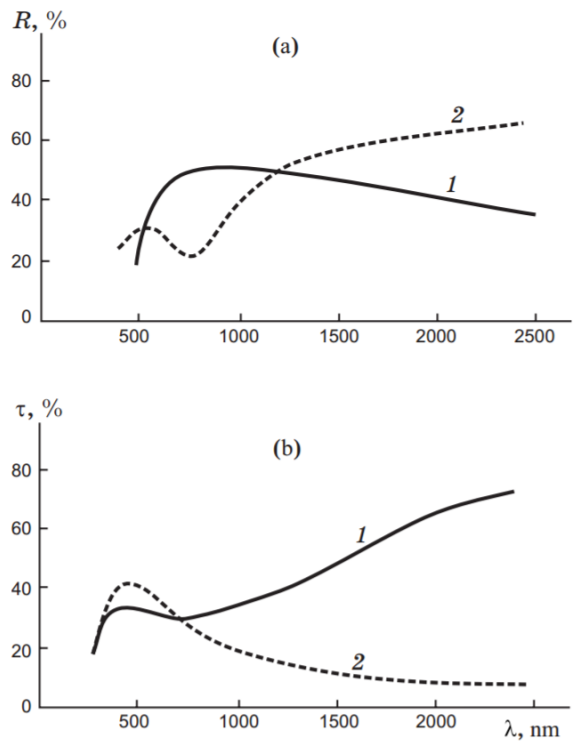

Figure 2: Reflection spectra (a) and transmission spectra (b) of the two samples referenced.

Several prototypes of linear bolometric detector arrays were fabricated. These arrays consist of 1 x 64, 1 x 128 and 1 x 256 pixels with lateral dimensions of either 50 pm x 50 pm or 100 pm x 100 pm. Fill factors for these detectors range from 25 % to 58 %, and linear array pitch is either 54 im (50 pm pixels) or 104 pm (100 pm pixels). Examples of the fabricated bolometric detector arrays are shown in Figs. 7 and 8 which is described by:

$$H = k \frac{\sigma T^4 A_s}{\pi d^2}$$

The basic parameters of the fabricated detectors such as responsivity (9), noise equivalent power (NEP), normalized detectivity (D*), and response time (t) were evaluated using the developed test bench schematically shown in Fig. 2. This test bench represents a typical measurement setup used for IR detector performance evaluation58. The JR source used consisted of a temperature controlled blackbody with a limiting aperture followed by a chopper. The blackbody temperature was kept at 500 K. The chopper, the current source and the lock-in amplifier were controlled by a computer. This allowed for automatic measurement of signal, noise and resistance, versus bias current or frequency. The constant k was determined using the conversion table of reference 8 which takes into account the dimension of the chopper teeth and slots. The obtained value is 0.443 which is close to the perfect square wave constant of 0.45. Parameters which are not taken into account in equation (2) are the radiation emitted by the chopper at 300 K and the atmospheric absorption. However these factors can be considered. The detectors were tested both in vacuum and at atmospheric pressure. The former type of measurements was performed placing the packaged detector chips in a custom-made vacuum chamber equipped with a germanium window. A limited transmittance of this window was taken into account in the performed measurements. A DC current source was used to bias the tested detector and the AC signal and noise at the chopper frequency were measured using a lock-in amplifier. A signal analyzer was used for measurements of noise spectrum. A low noise voltage amplifier and preamplifier were used to ensure that the measurement setup noises were lower than the tested bolometer noise. The noise of the preamplifier (Stanford Research System SR550) was 5.5nV/Hz112 at 30 Hz and its gain was 10. The SR560 amplifier had a 8 nV/Hz"2 noise and a gain of 100, which is shown in the equations:

$$NEP = \frac{V_n}{\mathfrak{R}}$$

$$D* = \frac{\sqrt{A_d \Delta f}}{NEP}$$

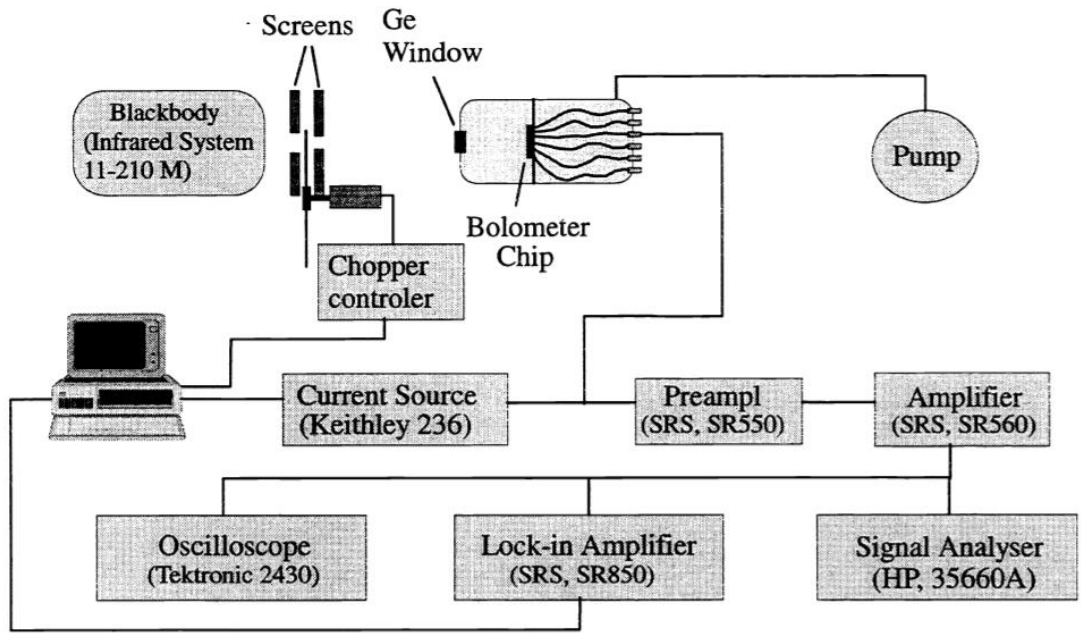

Figure 3: Schematic of the testbench used for the detector performance evaluation.

Typical parameters for the fabricated bolometric detectors measured in vacuum are listed in Tab. 1. The D* values obtained at atmospheric pressure were almost one order of magnitude lower than the values measured in vacuum. This underscores the importance of vacuum packaging of the bolometric detectors. Examples of the recorded characteristics for one of the test detectors (marked with an asterisk in Fig. 4) are shown in Fig. 1. Notice a negative slope of the voltage versus current characteristics (Fig. 1) for bias currents exceeding 30 j.tA. This region corresponds to the beginning of the semiconductor-to-metal phase transition in the v02 film due to Joule heating of this film by the bias current. The detector D* (Fig. l) also reaches its maximum at the onset of this phase transition while R (Fig. 3) does not vary significantly for the bias currents from about 15 .tA to about 40 .tA region. Detector responsivity depends strongly on frequency (Fig. l) decreasing from about 45,000 V/W at 2 Hz to about 10,000 V/W at 100 Hz. The detector response time slightly exceeds 7 ms (Figs. lOe andf).

An analysis of the optical characteristics of single layers of vanadium dioxide when they undergo a phase transition showed that their changes are very different in the visible, near- and mid-IR regions. As can be seen from the table, the strongest changes of optical constants n and k are observed in the spectral interval 9–11 m, where they take the maximum values in the metallic phase. However, it is obvious that single films of vanadium dioxide have a limited modulation depth in both reflected and in transmitted light. A numerical analysis of the possibilities of increasing the contrast of the reflectance changes of single layers of vanadium dioxide showed that the bistable properties can be improved by depositing them on highly reflective metallic mirrors.

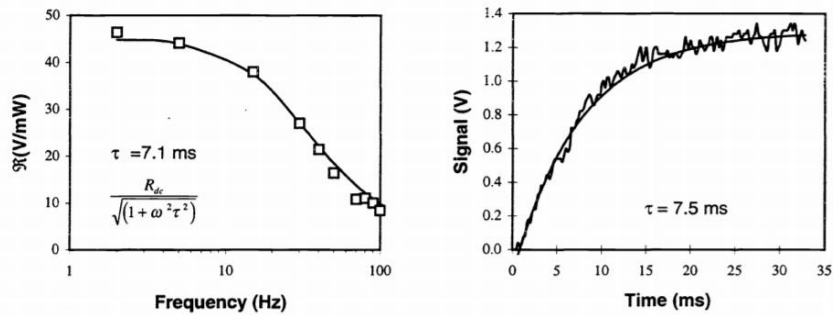

Figure 4: Voltage response with time and responsivity with frequency.

Because MEMS technology is relatively new, few computer simulation tools currently exist. In particular, no commercial software is available for simulating microbolometer performance. For this reason, a custom simulation software tool was developed, which provides a two-dimensional solution of the governing differential equations. The advantage of this approach is that it is more accurate than simple lumped models, and it inherently provides detailed heat flow and temperature distributions of the bolometer surface. These distributions can be a valuable aid in the design and optimization process. The developed software was designed to accommodate arbitrary bolometer geometries; an input file contains a description of the bolometer geometry, along with associated material parameters for the thin films which form the device. In addition, this software also accounts for the non-uniform absorption spectrum of the bolometer over the range of wavelengths from 2 im to 50 j.tm. For a given design, the software calculates the associated key figures of merit required to assess the bolometer performance. These figures of merit include 9t, NEP, D*, and t. Measurement and simulation values for responsivity versus bias current are shown in Fig. 4. Good matching of the theoretical and experimental results is clearly visible.

However, even in this case, it is extremely difficult to obtain films of a strictly specified composition because of the disproportioning of the partial pressure of the separate components of the composite material as a consequence of the differences of their evaporation temperatures. As a result of the indicated process, the differences in the bistable characteristics of the layers obtained by evaporating composite materials of different compositions were levelled off by other technological factors. Thus, films obtained when pure vanadium pentoxide was evaporated differed little in their characteristics from layers obtained when its composite with tungsten oxide was evaporated. As a result, a major role was played not by the starting composition of the film-forming material but by the conditions of the subsequent oxidative– reductive annealing. Even though the latter made it possible to fabricate VOx layers with satisfactory phase-transition characteristics, such a technology is rather laborious. The optical parameters of such layers make it impossible to count on obtaining bistable elements with fundamentally new properties by these methods. The further development of the technology for creating such elements is obviously associated with new methods for forming vanadium dioxide films. Methods for obtaining epitaxial films present the greatest interest in this case. Such films can be fabricated in a single process cycle by the method of magnetron reactive ion– plasma sputtering. It is possible to obtain films of composite composition by simultaneously sputtering two target cathodes co-sputtering. It is more realistic in our view to sputter from a single target cathode of composite composition. In this connection, the possibility should be investigated of fabricating such target cathodes up to 75 mm across and 4–5 mm thick. Going to an RF sputtering method makes it possible to eliminate the distortion of the congruence of the composition of the component being sputtered and the target, regardless of the melting temperature of the separate components.